# Dimensionality-driven orthorhombic MoTe$_2$ at room temperature

Rui He,[1,2,*] Shazhou Zhong,[3,4] Hyun Ho Kim,[3,5] Gaihua Ye,[2] Zhipeng Ye,[1] Logan Winford,[2] Daniel McHaffie,[3,5] Ivana Rilak,[3] Fangchu Chen,[6,7] Xuan Luo,[6] Yuping Sun,[6,8,9] and Adam W. Tsen[3,5*]

[1]*Department of Electrical and Computer Engineering, Texas Tech University, Lubbock, Texas 79409, USA*
[2]*Department of Physics, University of Northern Iowa, Cedar Falls, Iowa 50614, USA*
[3]*Institute for Quantum Computing, University of Waterloo, Waterloo, Ontario N2L 3G1, Canada*
[4]*Department of Physics, University of Waterloo, Waterloo, Ontario N2L 3G1, Canada*
[5]*Department of Chemistry, University of Waterloo, Waterloo, Ontario N2L 3G1, Canada*
[6]*Key Laboratory of Materials Physics, Institute of Solid State Physics, Chinese Academy of Sciences, Hefei 230031, People's Republic of China*
[7]*University of Science and Technology of China, Hefei, 230026, China*
[8]*High Magnetic Field Laboratory, Chinese Academy of Sciences, Hefei 230031, People's Republic of China*
[9]*Collaborative Innovation Centre of Advanced Microstructures, Nanjing University, Nanjing 210093, People's Republic of China*

[*]Correspondence to: rui.he@ttu.edu, awtsen@uwaterloo.ca

**Abstract:**

We use a combination of Raman spectroscopy and transport measurements to study thin flakes of the type-II Weyl semimetal candidate MoTe$_2$ protected from oxidation. In contrast to bulk crystals, which undergo a phase transition from monoclinic to the inversion symmetry breaking, orthorhombic phase below ~250 K, we find that in moderately thin samples below ~12 nm, a single orthorhombic phase exists up to and beyond room temperature. This could be due to the effect of *c*-axis confinement, which lowers the energy of an out-of-plane hole band and stabilizes the orthorhombic structure. Our results suggest that Weyl nodes, predicated upon inversion symmetry breaking, may be observed in thin MoTe$_2$ at room temperature.

**Main text:**

Layered transition metal dichalcogenides (TMDCs) are a rich family of compounds that crystallize in several different polytypes. While the semiconducting 2H structure has been studied extensively for electronic and optoelectronic device applications [1,2], the metallic 1T structure hosts various collective electron phases such as charge density waves and superconductivity [3]. MoTe$_2$ is one of the few TMDCs in that it stabilizes both semiconducting and metallic polytypes, transitions between which can be further controlled by temperature, alloying, strain, and electrostatic gating [4–10]. 1T-MoTe$_2$ is unstable, however—distortion of in-plane bonds gives rise to an enlarged monoclinic unit cell ($\beta$ or 1T' phase) at room temperature [11]. Below ~250 K, this structure changes further as a shift in layer stacking produces an orthorhombic crystal ($\gamma$ or T$_d$ phase) [12]. A schematic of this stacking distortion is shown in Fig. 1(a). This low temperature state exhibits a number of interesting properties, such as extremely large magnetoresistance [5,13,14], superconductivity with possible unconventional origins [15–17], and type-II Weyl nodes [18–24], the last of which requires broken inversion symmetry established only within the $\gamma$ phase.

Since the $\beta - \gamma$ phase transition involves an out-of-plane distortion, it may be possible to tune this transition by changing dimensionality. Here, we demonstrate using Raman and transport measurements that $\gamma$-MoTe$_2$ is observed in moderately thin samples below ~12 nm at temperatures up to 400 K. The mechanism may originate from the inherent three-dimensional (3D) band structure of MoTe$_2$—reducing thickness confines hole carriers along the *c*-axis, which stabilizes the orthorhombic phase in accordance with theoretical predictions [25]. Our results then suggest that thin MoTe$_2$ may exist as a Weyl semimetal at room temperature.

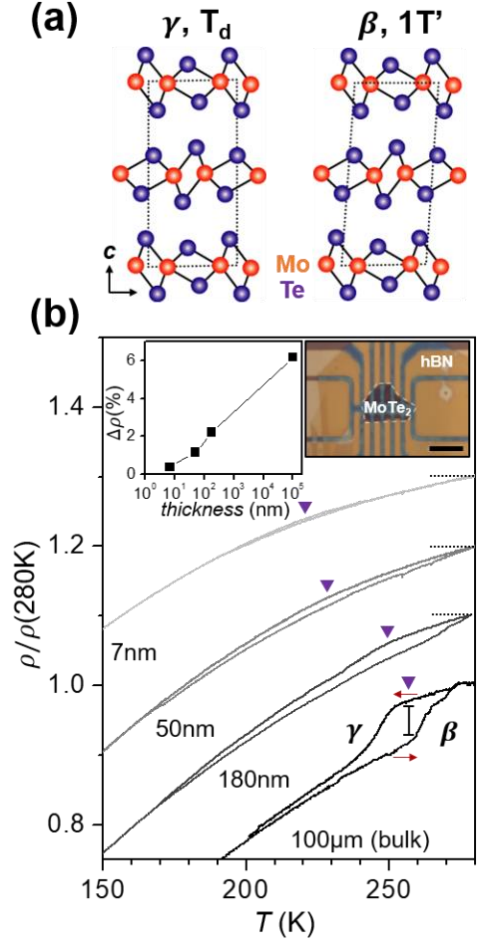

FIG. 1 (a) Structure of orthorhombic ($\gamma$ or $T_d$) and monoclinic ($\beta$ or 1T') phases of MoTe$_2$. (b) Right inset: optical image of thin flake device capped with hBN to protect from sample oxidation. MoTe$_2$ is outlined with dashed line. Scale bar is 10 µm. Main panel: normalized temperature dependent resistivity of MoTe$_2$ bulk crystal and thin flakes. Upper traces are offset for clarity. Kink and hysteresis between cooling and warming at 250 K corresponds to first-order $\beta - \gamma$ phase transition. Hysteresis becomes less visible in thinner flakes. Left inset: percentage resistivity difference between cooling and warming in the middle of hysteresis region (marked by purple arrows) as a function of flake thickness.

The upper right inset of Fig. 1(b) shows an optical image of a representative device with the MoTe$_2$ flake outlined by a dashed line. In order to avoid the effects of surface oxidation [26–28], MoTe$_2$ was exfoliated onto a polymer stamp within a nitrogen-filled glovebox, transferred onto gold electrodes, and covered with hexagonal boron nitride (hBN) before being moved out to the ambient environment. Since hBN conforms to the features of the underlying surface, an atomic force microscope was used to determine the thickness of the buried MoTe$_2$. The electrodes are etched into the oxidized silicon wafers to allow a more planarized surface for MoTe$_2$ transfer (see Supplemental Material). In main panel of Fig. 1(b), we show temperature dependent resistivity, normalized to the resistivity at 280 K, $\rho(T)/\rho(280 \text{ K})$, for 1.2K/min cooling and warming of three thin MoTe$_2$ samples (thickness 7, 50, and 180 nm) prepared in this way as well as that of a bulk crystal (thickness 100 µm). The traces for the thin flakes are offset vertically for clarity, and the offset values are marked by dashed lines on the right.

In contrast with an earlier study on unprotected MoTe$_2$ flakes, which reports a metal-to-insulator transition in samples below ~10 nm thickness [5], all samples here show metallic behavior down to 150 K, i.e. $\frac{d\rho}{dT} > 0$, with negligible backgate voltage dependence (see Supplementary Material). This indicates that insulating behavior is likely caused by surface degradation and is not intrinsic to reduced dimensionality. Second, while the resistivity of the bulk crystal shows a kink and a hysteresis between cooling and warming around ~250 K, as indicative of the $\beta - \gamma$ phase transition [29], the hysteresis becomes less apparent with decreasing thickness and is barely visible for the 7 nm flake. The hysteresis loop is also not sensitive to changing temperature sweep rates (see Supplementary Material). To quantify this trend, in the upper left inset of Fig. 1b, we plot the percentage resistivity difference between cooling and warming, $\Delta\rho$, measured in the middle of the hysteresis region (marked by purple arrows in the main panel) as a function of sample thickness in log scale. We observe that $\Delta\rho$ is substantially reduced from the bulk value even for a relatively large thickness of 180 nm, which is unexpected in that it contains over 250 layers (single layer thickness is ~0.7 nm) and reflects that the $\beta - \gamma$ phase transition in MoTe$_2$ is essentially 3D in character.

Results from the resistivity measurements presented above suggest one of two scenarios: for thin samples either 1) the $\beta - \gamma$ phase transition proceeds gradually with changing temperature, or 2) only a single phase exists throughout the entire temperature range. In order to discriminate between the two, we have performed temperature dependent Raman spectroscopy, which has been demonstrated to clearly distinguish between the $\beta$ and $\gamma$ phases of bulk MoTe$_2$ [30–32]. In our optical microscopy setup, we focused a linearly polarized, 532 nm wavelength laser through either a 50× (for samples in the cryostat) or a 100× (for samples outside cryostat) objective, yielding a ~2 µm or ~1 µm spot size, respectively. The use of successive notch filters allows measurement reaching down to ~5 cm$^{-1}$ shift from the laser line. Thin flake samples for Raman spectroscopy were also prepared in the glovebox on substrates without electrodes and covered with hBN.

For the top panel of Fig. 2(a), we first show Raman spectra for 50-nm- and 20-nm-thick flakes taken both at 294 K. A single peak at 128 cm$^{-1}$ is observed within the plotted range, similar to what has been observed previously for $\beta$-MoTe$_2$ bulk crystals [31,32]. For the upper trace in the lower panel, we show spectra for the 50-nm-flake at 140 K. Here, three peaks (at 12, 128, and 130 cm$^{-1}$) are seen, similar to what has been observed in bulk $\gamma$-MoTe$_2$ [31,32]. We have followed the convention used by Zhang *et al.* and labeled the $\gamma$ phase peaks as A, D, and E in order of increasing energy [32]. The lower energy A peak has been assigned to a shear mode, while the new higher energy peak is attributed to an out-of-plane vibrational mode, both of which become activated as a consequence of inversion symmetry breaking in the orthorhombic state [31,32].

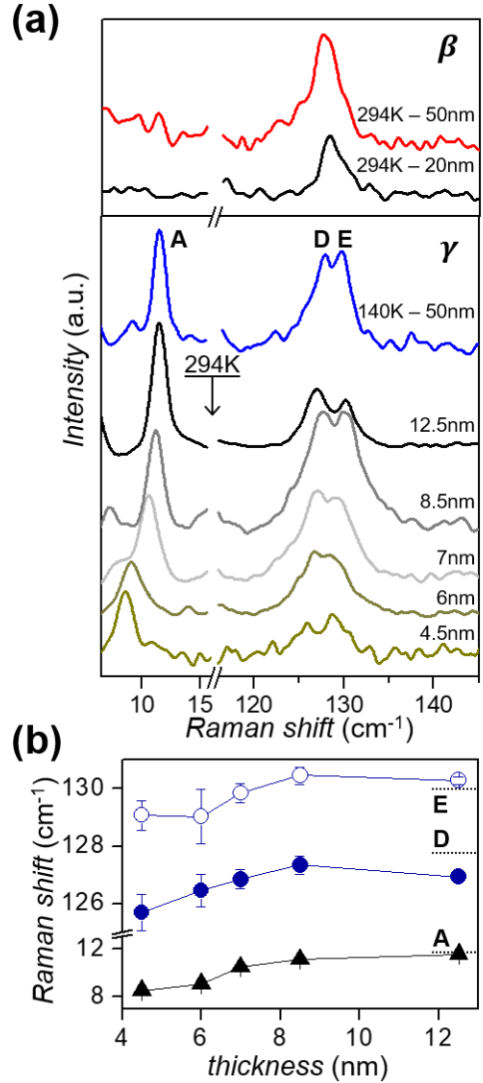

FIG. 2 (a) Main panel, top two traces: Raman spectra of 50 nm and 20 nm MoTe$_2$ at room temperature resembling $\beta$ phase. Lower traces: spectra of 50 nm MoTe$_2$ at 140 K ($\gamma$ phase) and thinner flakes at 294 K. $\gamma$-MoTe$_2$ shows additional peaks at ~12 cm$^{-1}$ and ~130 cm$^{-1}$, indicating thin samples are in $\gamma$ phase at room temperature. (b) Raman mode positions vs. thickness for $\gamma$-MoTe$_2$ flakes at 294 K. Corresponding mode positions for 50 nm sample at 140 K are marked by dashed lines.

For the lower traces, we compare Raman spectra for four thinner flakes of different thicknesses (4.5, 6, 7, 8.5, and 12.5 nm) taken at 294 K. Interestingly, they show similar features to bulk $\gamma$-MoTe$_2$ at low temperature and exhibit three peaks instead of one. In Fig. 2(b), we explicitly plot these peak positions as a function of flake thickness. We have also marked with dashed lines the energies of modes A, D, and E observed in 50 nm MoTe$_2$ within the $\gamma$ phase at 140 K. Overall, the thin flake modes redshift with decreasing sample thickness. Similar softening has been observed in other TMDC materials and could be due to a reduced interlayer force constant in few layer systems [33–35]. The extrapolation of these mode positions to the bulk $\gamma$ phase peaks in the thick limit, however, indicates that thin MoTe$_2$ ($\lesssim$12 nm) exhibits the inversion symmetry breaking orthorhombic structure at room temperature. We thus designate these three peaks as A, D, and E in direct connection with this phase. Since Weyl nodes appear in MoTe$_2$ as a consequence of inversion symmetry breaking, this observation suggests that thin MoTe$_2$ is already a Weyl semimetal at room temperature.

In order to confirm that the $\gamma$ phase is established across the entire measured temperature range for thin samples (scenario 2), it is necessary to perform Raman measurements with changing temperature. In the main panel of Fig. 3(a), we plot the evolution of Raman spectra for another 4.5-nm-thick sample upon both cooling to 150 K and warming to 400 K. The A, D, and E modes characteristic of the $\gamma$ phase can be seen at all temperatures. The additional shoulder seen at ~10 cm$^{-1}$ below room temperature is most likely an artifact (see Supplementary Material). We have fitted these three peaks with Lorentzian lineshapes and plotted their temperature dependent mode positions and areal intensities in Fig. 3(b). The modes redshift with increasing temperature, and increasingly so above room temperature. For cooling below 300 K, the intensity of peak E grows gradually with decreasing temperature, while the intensities of A and D slightly decrease. None of these modes, however, display large, abrupt changes characteristic of a first-order $\beta - \gamma$ phase transition as in the bulk crystal at ~250 K [31,32]. For heating close to 400 K, the intensities of all three peaks decrease, which could indicate the beginning of a transition into another phase different from both the $\beta$ and $\gamma$ phases.

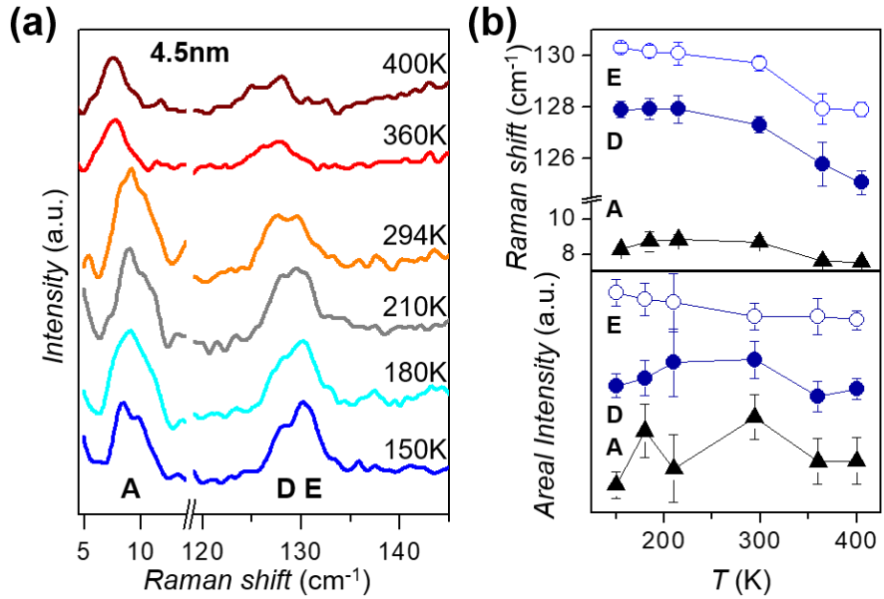

FIG. 3 (a) Temperature evolution of Raman spectra for 4.5 nm sample upon cooling and heating from room temperature. (b) Mode positions and areal intensities vs. temperature after Lorentzian fits. Traces have been offset for clarity. No abrupt changes at ~250 K corresponding to first-order $\beta - \gamma$ phase transition is observed.

The combined results of our Raman and transport study can be summarized by the temperature-thickness phase diagram shown in Fig. 4(a). For samples 12.5 nm and thinner, the $\gamma$ phase is stabilized at all temperatures up to 400 K. For thicker samples, a phase boundary separates the low-temperature $\gamma$ phase from another high temperature phase, which in the limit of a bulk crystal is the $\beta$ phase. We have used the center temperature of the resistivity hysteresis to mark this phase boundary down to 50 nm (see Fig. 1(b)). Twenty nanometers is the smallest thickness measured for which the $\gamma$ phase is not observed in Raman at room temperature (see Fig. 2(a)). We have cooled this sample and determined that it transitions to $\gamma$ phase at ~210 K. This indicates that transition temperature doesn't change substantially with decreasing thickness, but rather terminates abruptly for a critical thickness (between 12.5 and 20 nm), below which only a single $\gamma$ phase exists.

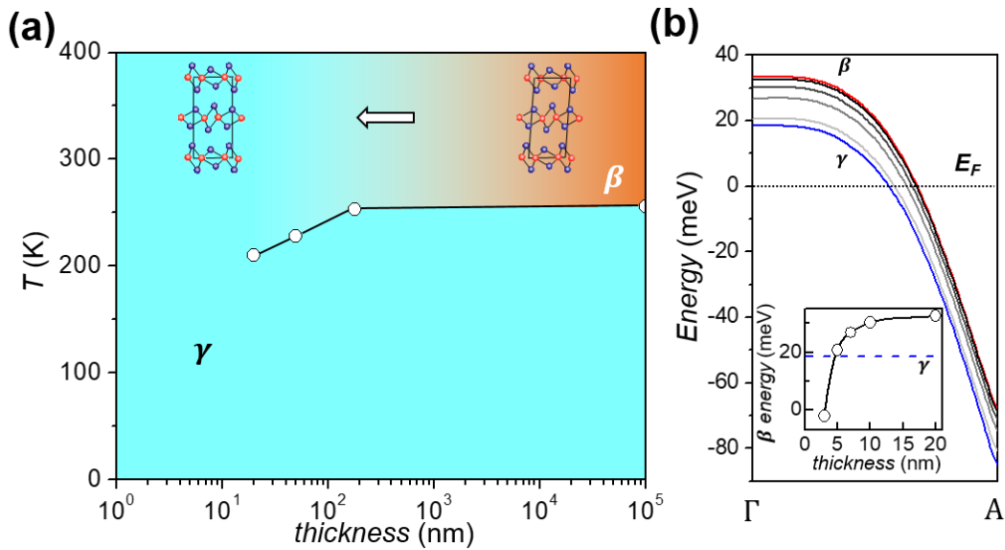

FIG. 4 (a) Temperature-thickness phase diagram. Below critical thickness a single $\gamma$ phase is stabilized at all temperatures up to 400 K. The high temperature $\beta$ phase undergoes crossover with reducing thickness, likely due to slowly changing structure. (b) Main panel: proposed mechanism for crossover. An out-of-plane hole band is shifted to lower energy upon cooling in thick samples and confinement in thin samples. Red and blue traces are reproduced from Kim *et al.* (ref. 25) and correspond to bulk MoTe$_2$ in the $\beta$ and $\gamma$ phase, respectively. Gray traces correspond to the $\beta$ band confined to thickness 20, 10, 7, and 5 nm. Inset: $\beta$ band energy vs. thickness at the $\Gamma$ point, showing crossing below the $\gamma$ phase energy (dashed blue line) at ~5 nm.

What is the nature of the high temperature phase? If we take the resistivity difference in the hysteresis region as an indicator of the difference between the high and low temperature states, its decrease with decreasing thickness (see Fig. 1(b) inset) indicates that the $\beta$ phase undergoes a slow crossover as thickness is reduced from the bulk limit. Since the $\beta - \gamma$ transition involves only slight tilt in the angle of the unit cell (~4°), this angle likely changes gradually until the $\gamma$ phase is reached below the critical thickness. This is illustrated in Fig. 4(a) by the fading colors.

We now examine the possible mechanisms for stabilization of $\gamma$-MoTe$_2$ in thin samples. First, since the dimensional crossover begins for thicknesses over 100 nm, it is unlikely that the cause is surface or substrate effects impacting the top- or bottom-most layers [28,36]. Consistent with this, changing the protecting layer from hBN to graphite does not alter the transition (see Supplementary Material). Second, since the Raman peak positions of thin $\gamma$-MoTe$_2$ are overall very similar to their bulk counterparts, especially for slightly thicker flakes close to 10 nm, we can also rule out the possibility of changing interlayer interactions driving the crossover [33–35]. Instead, we look to the 3D origins of the $\beta - \gamma$ transition in bulk systems.

Recently, Kim *et al.* has calculated the electronic structure of bulk MoTe$_2$ across this phase transition [25]. Starting in $\beta$ phase, they find that there are two hole pockets that cross the Fermi energy in the out-of-plane, Γ to A direction. In addition to changes in the in-plane band structure, the upper *c*-axis hole band shifts to lower energy upon transition into the $\gamma$ phase. In comparison, for WTe$_2$, a structurally similar compound which always exists in the $\gamma$ phase, the corresponding hole bands sit below the Fermi level. As a consequence, Kim *et al.* predicts that $\gamma$-MoTe$_2$ can be stabilized by electron doping.

Quantum confinement produced by thickness reduction may yield a similar stabilization of the orthorhombic $\gamma$ phase by pushing the hole bands to lower energy. In the main panel of Fig. 4b, we have used the results of Kim *et al.* and added a *c*-axis confinement energy, $\Delta E = \frac{\hbar^2 \pi^2}{2m_\perp L^2}$, to the $\beta$ hole band for various thicknesses $L$ (20, 10, 7 and 5 nm) by evaluating the effective mass through a numerical derivative ($m_\perp \sim 1.18 m_e$). Comparing to the corresponding band in the $\gamma$ phase, we observe that thickness reduction results in a continuous shift of the $\beta$ phase band towards that of the $\gamma$ phase. In particular, as shown in the inset, the confined $\beta$ band energy at the $\Gamma$ point crosses the $\gamma$ band for thicknesses under ~5 nm. This critical thickness is less than but within the same order of the experimentally determined value, which could be due to an overestimate of the true effective mass and/or suggests that it may be necessary to consider the complete Fermi surface in order to fully substantiate this scenario. This simple analysis nevertheless provides for both a reasonable estimate of the critical thickness and accounts for the changes we observe in samples beyond the few-layer limit. An analogy may also be made to the semiconducting TMDCs, such as $MoS_2$ and $WSe_2$. For these materials, since the out-of-plane effective mass for both electrons and holes at the K point is much larger than those closer to the $\Gamma$ point, decreasing thickness leads to a large confinement energy increase for the indirect gap, while the direct gap is relatively unchanged [37].

In summary, we observe a single orthorhombic phase up to 400 K in $MoTe_2$ flakes thinner than ~12 nm. Thicker samples may exist in a transitional state between the $\beta$ and $\gamma$ structures—further investigations are needed to explore this. The orthorhombic phase breaks inversion symmetry and is expected to yield type-II Weyl nodes in the band structure. Amongst different possible causes, we considered the effect of perpendicular confinement on out-of-plane hole bands. We expect that this shift will have measurable effects on the transport properties of thin samples, especially at low temperatures where the electron and hole concentrations are balanced in the bulk crystal.

**Acknowledgements:**

We thank Anton Burkov and Walter Lambrecht for helpful discussions. RH, ZY, and LW acknowledge support by the National Science Foundation (NSF) CAREER Grant (No. DMR-1760668). GY acknowledges support by the NSF RUI Grant (No. DMR-1410496). The low temperature equipment was acquired through the NSF MRI Grant (No. DMR-1337207). AWT acknowledges support from an NSERC Discovery grant (RGPIN-2017-03815). This research was undertaken thanks in part to funding from the Canada First Research Excellence Fund. Work in China was supported by the National Key Research and Development Program under contracts 2016YFA0300404 and the National Nature Science Foundation of China under contracts 11674326 and the Joint Funds of the National Natural Science Foundation of China and the Chinese Academy of Sciences' Large-Scale Scientific Facility under contracts U1432139.